**Explainable Self-Organizing Artificial Intelligence Captures Landscape Changes Correlated with Human Impact Data**

1 John Mwangi Wandeto, 2 Birgitta Dresp-Langley*

1 Department of Information Technology, Dedan Kimathi University of Technology, Nyeri, Kenya, john.wandeto@dkut.ac.ke

2 UMR 7357, Centre National de la Recherche Scientifique (CNRS), Strasbourg University, France,

*birgitta.dresp@cnrs.fr

*Corresponding author: Birgitta Dresp-Langley, UMR 7357 CNRS-Strasbourg University, France Email: birgitta.dresp@cnrs.fr

**Abstract:** Novel methods of analysis are needed to help advance our understanding of the intricate interplay between landscape changes, population dynamics, and sustainable development. Self-organized machine learning has been highly successful in the analysis of visual data the human expert eye may not be able to see. Thus, subtle but significant changes in fine visual detail in images relating to trending alterations in natural or urban landscapes, for example, may remain undetected. In the course of time, such changes may be the cause or the consequence of measurable human impact, or climate change. Capturing such change in time series of satellite images before the human eye can detect the signs thereof makes important trend information readily available at early stages to citizens, professionals and policymakers. This promotes change awareness, and facilitates early decision making for action. Here, we use unsupervised Artificial Intelligence (AI) that exploits principles of self-organized biological visual learning for the analysis of time series of satellite images. The Quantization Error (QE) in the output of a Self-Organizing Map prototype is exploited as a computational metric of variability and change. Given the proven sensitivity of this neural network metric to the intensity and polarity of image pixel contrast, and its proven selectivity to pixel colour, it is shown to capture critical changes in urban landscapes. This is achieved here on the example of satellite images from two regions of geographic interest in Las Vegas County, Nevada, USA across the years 1984-2008. The SOM analysis is combined with the statistical analysis of demographic data revealing human impacts significantly correlated with the structural changes in the specific regions of interest. By correlating the impact of human activities with the structural evolution of urban environments we further expand SOM analysis as a parsimonious and reliable AI approach to the rapid detection of human footprint-related environmental change.

## 1. Introduction

Rapid modification of human environments results in unprecedented environmental challenges that impact sustainability to greater or lesser extents depending on the specific geographic and social context. The study of urban landscape evolution and its correlation with human impact and demographic data helps understand its implications for the development of human activities in these urban ecosystems and their sustainability given the particular geographic context. Novel approaches and methods of analysis are needed to help advance our understanding of the intricate interplay between landscape changes, population dynamics, and sustainable development (Craglia and Nativi, 2018) under the light of a more or less limited availability of natural resources. Subtle changes in changes in earth imaging data may reflect the cause or the consequence of measurable human impact or climate change (Kotchi et al, 2019, Furusawa et al, 2023). Predicting or mapping the risks associated with these by exploiting artificial intelligence (AI) for the analysis of satellite images is an emerging method that can support research, surveillance, prevention and control activities (Sudmanns et al, 2019). This often implies resorting to methods of data analysis from several disciplines to address the problems of human impact on landscapes (Schipper, Dubash, and Mulugetta, 2021). Recognition of the complexity of the links between human impact data, landscape alterations, and climate change compels researchers to draw on interdisciplinary knowledge that marries the physical and natural sciences with social sciences and humanities. Given the slow unfolding of what may become catastrophic changes to human environments (Palmer and Stevens, 2019, Olsavszky et al, 2020), the methods of earth observation image analysis need to be sufficiently sensitive to the smallest detectable changes to permit risk-aware monitoring at different scales of time and space. Such highly sensitive methods then can inform the quantitative estimation and prediction of risks as well as risk mapping. The current array of Earth observation satellites provides access to a large quantity and variety of data with the highest possible spatial and temporal resolution (Kotchi et al, 2019), which facilitates their subsequent analysis.

There are various approaches to the analysis of Earth image data from satellites (Ahn et al, 2023, Kumar and Miklavcic, 2018, Metzler et al, 2023, Dou et al, 2023, Luke, 2013) to study temporal changes in landscape data that may represent meaningful phenomena in terms of human impact (Camacho et al, 2014) or climate change (Orheim and Lucchitta, 1990, Bhaskar et al, 2010). Here, we use a pixel color-based approach to the analysis of Earth imaging data to show that the method consistently captures the earliest signs of structural change in satellite images of two specific geographical regions of interest (ROI) across a time period where critical landscape changes occurred. These changes are further shown to be significantly correlated with human data and demographics from public archives for the same time period, highlighting critical dimensions of human impact on the expansion of built environments and the stresses this exerts on survival relevant natural resources (Figure 1). NASA Landsat images (NASA Goddard Space Flight Center, 2012) of Las Vegas City Centre and the residential North of Las Vegas from the time period between 1984 and 2008 were submitted to analysis for this study. In the 1980ies Las Vegas City, located in the middle of the Nevada Desert, featured mainly 'The Strip', with a number of smaller casinos and motels. Subsequently, in a large restructuration project between 1990 and 2007 a large number of the old casinos and motels were demolished. The ensuing reconstruction of Las Vegas City Centre and the subsequent opening of a large number of mega-resorts with casino spaces, artificial tropical landscapes and urban large-scale simulations, restaurants with world-class chefs, and shows performed by international megastars transformed Las Vegas City Centre and The Strip entirely. By offering multiple kinds of entertainment, dining gambling, and lodging, attracting millions of visitors (Las Vegas Convention and Visitors Authority, 2024) from all over the world, Las Vegas City Centre has become one of the largest entertainment poles in the world (Luke, 2013). Most elements of this project opened in late 2009. This was accompanied by the rapid spread of greater Las Vegas, including the residential North, into the adjacent desert. The population count (Las Vegas Population Review, 2024) grew from thousands in 1984 to millions in 2009. Given the specific geographic location and the limited natural resources therein, the expanded urbanization of Las Vegas had its consequences on its own sustainability, as demonstrated in a recent study suggesting that conservation and other strategic water supply measures now have become urgently necessary to ensure water supply for Las Vegas in the future (Dow et al., 2019, US Department of Interior Bureau of Reclamation Hoover Dam Control Room Statistics, 2023).

**Figure 1**
Consequences of human activity on urban development and natural resources

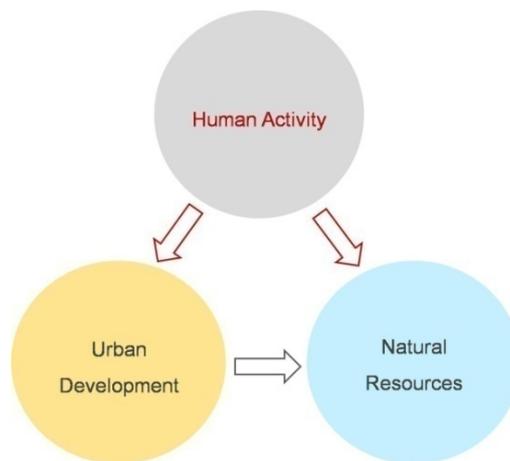

Although the current array of Earth observation satellites such as NASA's Landsat provides access to a large quantity and variety of data with the highest possible spatial and temporal resolution (Kotchi et al, 2019), any technique for studying changes in imaging data across time sooner or later requires choosing a method and a model for an interpretation. This introduces potential sources of bias (Tan, Gao, Khan, and Guan, 2022, Saint James Aquino, 2023, Sylolypavan et al, 2023), even when the image analysis is combined with an advanced imaging model, which itself has to be explainable. In computer and AI-assisted methods for the analysis of image time series, there are various potential sources of bias (Brown, 2017, Baxter and Jannin, 2022) at essentially three levels:

1) input/training data and/or their configuration (MacNamee et al, 2002, Daneshjou et al, 2021, Gichoya et al, 2023)

2) learning algorithms (Wieland and Pittore, 2014, Nazer et al, 2023)

3) interpretation of output data (Brown, 2017, Drukker et al, 2023)

Also, a meaningful interpretation of the data may involve guesswork when the uncertainty about trends or correlations is high. In short, it is difficult to rule out subjectivity of the human agent at various steps in the process, even when the latter is a skilled expert. In addition, sustainable technology for data classification needs to combine high accuracy with further advantages relative to economy in resource demands, computing speed, objectivity, and output reproducibility. The proposed approach ensures these criteria (Wandeto and Dresp-Langley, 2023), and excludes human bias at any of the three levels mentioned above. The method consists of exploiting the quantization error (QE) in the output of a SOM (Kohonen, 1998, 2001, 2014), a self-organizing neural network architecture that maps pixel color input (RGB) from large image data, i.e. images containing up to several millions of pixels, with the highest possible (to the single pixel) precision on the basis of unsupervised, competitive winner-take-all learning (Carpenter, 1997, Kohonen, 1998, Yang and Chen, 2000, Chen, 2017). The QE in the SOM output, frequently used as a quality metric for network quantization (Kohonen, 2001, deBodt, Cottrell, and Verleysen, 2002, Castagnetti, Pegatoquet, and Miramond, 2023), has more recently proven a consistent detector of invisible local changes in streams of thousands or millions of, randomly or systematically varying, input data (Wandeto and Dresp-Langley, 2019 a, b, c, Wandeto and Dresp-Langley, 2023). When the SOM input is constant across time, the quantization error is invariant; when variations in a locally defined dimension of the input data occur across time, the QE will reliably increase or decrease depending on the direction of change. This novel aspect of input-driven self-organization (Zhang, 1991, Grossberg, 1993, Ligaya and Fusi, 2013 Eglen and Gjorgjieva, 2009) can be, as is shown here, strategically exploited in artificial neural network maps with a minimalistic functional design where the model neurons become locally and globally ordered during learning (Erwin, Obermayer, and Schulten, 1992; Ota et al, 2011). Unsupervised winner-take-all learning in SOM is akin to biological synaptic learning (Hebb, 1949, Grossberg, 1993,) and the self-organizing functional properties of biological sensory neurons (e.g. Buonamano and Merzenich, 1998). The output metric as defined provides a highly reliable indicator that scales, in a few minutes and with a to-the-single-pixel precision, local variability in time series of images containing millions of pixels each. This was previously demonstrated on image time series of natural environments (Wandeto, Dresp-Langley, and Nyongesa, 2018) organs (Wandeto, Nyongesa, and Dresp-Langley, 2017; Wandeto, Nyongesa, Remond, and Dresp-Langley, 2017), cells (Wandeto and Dresp-Langley, 2019a, Dresp-Langley and Wandeto, 2020, 2021b), simulated visual objects (Dresp-Langley and Wandeto, 2021a) and temporal series of sensor data (Dresp-Langley, Liu, and Wandeto, 2021, Liu, Wandeto, Nageotte, Zanne, de Mathelin, and Dresp-Langley, 2023).

## 2. Theoretical Framework

The Self-Organizing Map (SOM) may be described formally as a nonlinear, ordered, smooth mapping of high-dimensional input data onto the elements of a regular, low-dimensional array (Kohonen, 1998, 2001, 2014). The general, fully connected functional architecture of the prototype for this study here is graphically represented in Fig. 2. It is assumed that the set of input variables can be defined as a real vector $x$, of n-dimensionality. A parametric real vector $m_i$ of n-dimension is associated with each element in the SOM. Vector $m_i$ is a model and the SOM is therefore an array of models. Assuming a general distance measure between $x$ and $m_i$ denoted by $d(x, m_i)$, the map of an input vector $x$ on the SOM array is defined as the array element $m_c$ that matches best (smallest $d(x, m_i)$) with $x$. During the learning process, the input vector $x$ is compared with all the $m_i$ in order to identify $m_c$. The Euclidean distances $\|x-m_i\|$ define $m_c$. Models topographically close in the map up to a certain geometric distance indicated by $h_{ci}$ will activate each other to learn something from their common input $x$.

**Figure 2**
SOM (cf. Kohonen, 1998) prototype with sixteen model neurons in a fully connected functional architecture

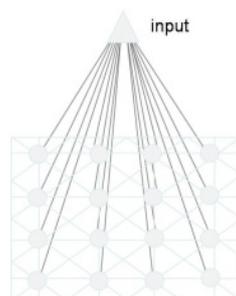

This results in a local relaxation or smoothing effect on the models in this neighborhood, which in continuous learning leads to global ordering. Self-organized learning in a SOM is represented by the equation

$$m(t+1)=m_i(t)+\alpha(t)\,h_{ci}(t)[x(t)-m_i(t)] \quad (1)$$

where t=1,2,3...is an integer, the discrete-time coordinate, hci (t) is the neighborhood function, a smoothing kernel defined over the map points which converges towards zero with time, $\alpha(t)$is the learning rate, which also converges towards zero with time and affects the amount of learning in each model. At the end of the winner-take-all learning process in the SOM, each image input vector x becomes associated to its best matching model on the map mc. The difference between x and mc, ||x-mc||, is a measure of how close the final SOM value is to the original input value and is reflected by the quantization error QE. The average QE of all Xi in an image is given by

$$QE= 1/N \sum_{(i=1)}^{N} | X_i-m_{(c_i)} | \quad (2)$$

where N is the number of input vectors x in the image. The final weights of the SOM here are defined by a three dimensional output vector space representing R, G, and B channels. The magnitude as well as the direction of change in any of these from one image to another is reliably reflected by changes in the QE. The spatial location, or coordinates, of each of the model neurons, or domains, placed at different locations on the map, exhibit characteristics that make each one different from all the others. When a new input signal is presented to the map, the models compete and the winner will be the model the features of which most closely resemble those of the input signal. The input signal will thus be classified or grouped in one of models. Each model or domain acts like a separate decoder for the same input, i.e. independently interprets the information carried by a new input. The input is represented as a mathematical vector of the same format as that of the models in the map. Therefore, it is the presence or absence of an active response at a specific map location and not so much the exact input-output signal transformation, or magnitude of response, that generates the interpretation of the input. To obtain the map size needed for a given study, a trial-and-error process is generally implemented (Kohonen, 1998, de Bodt, Cottrell, and Verleysen, 2002), which is directly determined by the value of the resulting quantization error; the lower this value, the better the "first guess".

## 3. Materials and Methods

### 3.1. Images

Earth observation satellites such as NASA's Landsat provide access to data with the highest possible spatial and temporal resolution (Kotchi et al, 2019). For this study, time series of satellite images NASA Landsat images (NASA Goddard Space Flight Center, 2012) of Las Vegas City Centre and the residential North of Las Vegas were submitted to analysis. The original images 25 images for each geographical ROI were extracted from time-lapse animations of Las Vegas County, Nevada, for a reference time period from 1984 to 2008, as captured by NASA Landsat sensors. VLC (2023), an open source media player, was used to generate static images from the time-lapse animations provided. The original images are false-colour, displaying arid desert regions in greenish-brown, building structures in gray levels and healthy vegetation/green spaces in red. Water is represented by black pixels. The 50 original static images can be downloaded from researchgate as indicated in the Data Availability Statement. Low-resolution copies of two of the high-resolution 25 images for the residential North from the years 1984 and 2008 are shown in Figure 3 for illustration.

**Figure 3**
Satellite image examples showing the residential North of Las Vegas in 1984 and 2008

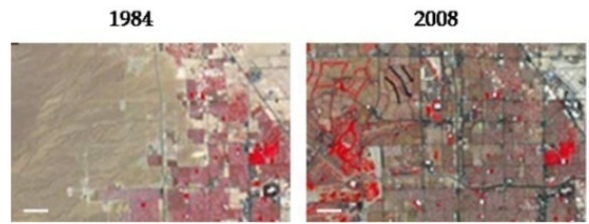

All images were pre-processed (Wieland and Pittore, 2014) to ensure they were identically scaled and aligned within a given series. This was achieved here through StackReg, a software tool for image co-registration specifically designed for scientific multidimensional image processing (Thévenaz, Ruttimann, and Unser, 1998) and the ancillary plugin TurboReg. Both are plugins to ImageJ, an open source image processing software package. ImageJ's script editor supports various programming languages including python and matlab. The co-registration method exploits an automatic subpixel registration algorithm minimizing the mean square intensity difference between a reference and a test data set based on a coarse to fine iterative strategy performed according to a new variation of the Marquardt-Levenberg algorithm for nonlinear least-square optimization (Thévenaz, Ruttimann, and Unser, 1998). StackReg is used as a front-end to TurboReg. Several mechanisms are at work to exchange data between these plugins, one of them involving temporary files which are written into the temporary directory that ImageJ defines. The StackReg software is available at http://bigwww.epfl.ch/thevenaz/stackreg/, with a direct link to TurboReg. The last image of the time series was used to anchor the registration. Control for variations in contrast intensity between images of a given series was performed after registration by increasing the image contrast and by removing strictly local variations at different times of image acquisition. For each extracted image, contrast intensity (I) normalization was ensured using

Ifinal = (I - Imin/Imax - Imin) x 255          (3)

The registered and normalized image taken in 2008, the last year of a time series from each ROI, was used to train the SOM for SOM-QE analysis.

## 3.2. SOM-QE analysis

Pixel RGB values were used as input stream to a four-by-four SOM with sixteen model neurons (Figure 2). This input dimensionality ensures the processing of fine image detail and permits avoiding errors due to inaccurate feature calculation, which occur frequently with complex images (Schneider, Rasband, and Eliceiri, 2012). To yield quantitative criteria for choosing the map size, a trial-and-error process (Kohonen, 1998, de Bodt, Cottrell, and Verleysen, 2002) was implemented using the image input data. This process led to observe that map sizes with more than 16 model neurons produced observations where some models ended up empty, which meant that these models did not attract any input towards the end of the training. As a consequence, a four-by-four SOM architecture with 16 models was sufficient to represent all the image data. Neighborhood distance and learning rate were constant at 1.2 and 0.2 respectively. These parameters arise on the basis of the same initial trial-and-error process when testing the quality of the "map's best first guess" (Kohonen, 1998, 2001, 2014), which is directly determined by the value of the resulting quantization error; the lower this value, the better the first guess. The models were initialized by randomly picking vectors from the training image, which allows the SOM to work on the original data without any prior assumptions about any level of organization within the data, but requires starting with a wider neighborhood function and a higher learning-rate factor than in procedures where initial values for model vectors are pre-selected. As a consequence, the SOM training here consisted of 1000 iterations for a two-dimensional rectangular map of 4 by 4 nodes capable of creating 16 model observations from the data. The spatial locations, or coordinates, of each of the 16 models or domains, placed at different locations on the map, exhibit characteristics that make each one different from all the others. When a new input signal is presented to the map, the models compete and the winner will be the model the features of which most closely resemble those of the input signal. The input signals are classified or grouped in one of models. Each model or domain acts like a separate decoder for the same input, i.e. independently interprets the information carried by new input. After the training phase, the QE in the trained map output was determined for each of the 25 images of the series. The code for running the SOM is written in Python 3 with numpy 1.24.2, available at https://github.com/SoftwareImpacts/SIMPAC-2023-308 in a special open access collection featuring software that has been verified for computational reproducibility by code ocean: https://codeocean.com/capsule/3541740/tree/v2.

## 4. Results

The results from the analyses on the image time series for the two geographical ROI show a general trend towards increase in the QE metric for each ROI between 1984 and 2008. These results are given in Table 1 as a function of the image year and the ROI.

**Table 1**
The quantization error in the SOM output (SOM-QE) as a function of the image year and geographic ROI

| Year | Las Vegas City | Residential North |
|------|----------------|-------------------|
| 1984 | 0,240437503 | 0,151226618 |
| 1985 | 0,264341069 | 0,157865360 |
| 1986 | 0,271480118 | 0,155998180 |
| 1987 | 0,289065099 | 0,169213765 |
| 1988 | 0,282803632 | 0,210600120 |

| 1989 | 0,286535270 | 0,213982186 |
| 1991 | 0,303956828 | 0,219973707 |
| 1991 | 0,298541690 | 0,225406972 |
| 1992 | 0,301994751 | 0,214975264 |
| 1993 | 0,293683986 | 0,208605453 |
| 1994 | 0,304328745 | 0,221313177 |
| 1995 | 0,298240329 | 0,212630331 |
| 1996 | 0,309779114 | 0,222464495 |
| 1997 | 0,284870821 | 0,225816809 |
| 1998 | 0,291493024 | 0,223026329 |
| 1999 | 0,296067339 | 0,238731572 |
| 2000 | 0,304491317 | 0,246826836 |
| 2001 | 0,311488540 | 0,254509105 |
| 2002 | 0,314104190 | 0,259835794 |
| 2003 | 0,299101833 | 0,263285485 |
| 2004 | 0,299139369 | 0,249477866 |
| 2005 | 0,296761075 | 0,237934169 |
| 2006 | 0,305053585 | 0,250044464 |
| 2007 | 0,301298833 | 0,256201407 |
| 2008 | 0,314321877 | 0,261825498 |

The numerical data were submitted to linear regression analysis to assess the statistical significance of the increase in SOM- QE across the years. The linear fits are shown in Figure 4 for each of the two ROI. As an estimate of the part of variance in the data that is accounted for by the linear trend, or fit, the regression coefficient r2 is computed. It is a direct reflection of the goodness of the fit. The statistical significance of the trend in the data in any given direction, upward or downward, is determined by the probability that the linear adjustment sufficiently differs from zero on the basis of Student's distribution (t). These results reveal a statistically significant linear trend towards increase in The SOM-QE as a function of time for each geographical ROI. The results from the linear regression analysis with the slopes and intercepts of the fits and their regression coefficients r2 are shown in Table 2. The results from the statistical trend analyses with Student's t, degrees of freedom (df) for a given comparison and the associated probability (p), limits are given in Table 3. The regression coefficients (Table 2) reveal that the linear fit to the SOM-QE for the images of the residential North is a reasonably good one, while the linear fit to the SOM-QE for Las Vegas City is not. This is consistent with the type of structural change that took place in each of the ROI across the study years. While there was a disruptive reorganization of the City Center, with old casinos and hotel centers disappearing to be replaced or not by new ones over the years, the landscape of what is now the residential North changed in terms of a much smoother and rather progressive, step-by-step development of building and housing agglomerations in these arid regions of the Nevada desert.

**Figure 4**
Linear fits to SOM-QE trends as a function of the image year and ROI

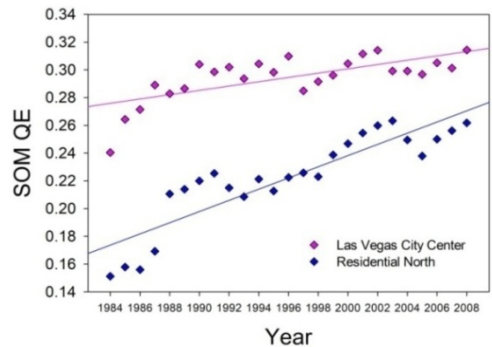

**Table 2**
Goodness of the linear fits to SOM-QE distributions for each geographical ROI

| Fit Parameter | Las Vegas City | Residential North |
|---|---|---|
| Slope ($b_1$) | 1,554 | 4,0331 |
| Intercept ($b_0$) | -2,8077 | -7,8286 |
| $r^2$ | 0,4776 | 0,7995 |

**Table 3**
Linear regression statistics with Student's t, degrees of freedom (df) and probability limits (p) for each geographical ROI

| Linear Regression | Las Vegas City | Residential North |
|---|---|---|
| t | 88,98 | 33,45 |
| df | (1, 24) | (1, 24) |
| *p* | <.001 | <.001 |

The linear trend statistics (Table 3) reveal a statistically highly significant increase in the SOM-QE across the image years for both ROI. For the same reference time period, the Las Vegas Convention and Visitors Authority (2024), and the Las Vegas Population Review (2024) have provided publically archived data that show a steep increase of human impact across the same years as those from which the satellite images analyzed here were taken. These data are shown in Table 4 in terms of annual population estimates in thousands for Greater Las Vegas, which includes the City and the residential North, and visitors in millions.

**Table 4**
Visitor and population statistics for Greater Las Vegas across the years of the study period

| Year | Visitors (in millions) | Population (in K) |
|---|---|---|
| 1984 | 12,8000 | 191,0000 |
| 1985 | 14,2000 | 197,0000 |
| 1986 | 15,2000 | 204,0000 |
| 1987 | 16,2000 | 226,0000 |
| 1988 | 17,2000 | 240,0000 |
| 1989 | 18,1000 | 266,0000 |
| 1991 | 20,9000 | 276,0000 |
| 1991 | 21,3000 | 298,0000 |
| 1992 | 21,9000 | 310,0000 |
| 1993 | 23,5000 | 330,0000 |
| 1994 | 28,2000 | 352,0000 |
| 1995 | 29,0000 | 374,0000 |
| 1996 | 29,6000 | 406,0000 |
| 1997 | 30,4000 | 423,0000 |
| 1998 | 30,6000 | 448,0000 |
| 1999 | 33,8000 | 466,0000 |
| 2000 | 35,8000 | 483,0000 |
| 2001 | 35,0000 | 506,0000 |
| 2002 | 35,0000 | 521,0000 |
| 2003 | 35,5000 | 535,0000 |
| 2004 | 37,3000 | 560,0000 |
| 2005 | 38,6000 | 576,0000 |
| 2006 | 38,3000 | 592,0000 |
| 2007 | 37,5000 | 603,0000 |
| 2008 | 39,5000 | 608,0000 |

These demographic data were also submitted to linear regression and statistical trend analysis. The linear trends are shown graphically in Figure 5. The results reveal statistically significant linear trends towards increase as a function of time for both types of human impact data. The results from the linear regression analysis with the slopes and intercepts of these fits and their regression coefficients r2 are shown in Table 5. The results from the statistical trend analyses with Student's t, Degrees of Freedom (df) for a given comparison and the associated probability (p), limits are given in Table 6. The regression coefficients r2 in Table 5 reveal that the quality of the linear fits to the human impact data trends in terms of annual visitor increase and population growth across the study years is excellent. The steady increase in population and visitors of Greater Las Vegas is consistent with the restructuration that took place across these years, creating an increasingly larger offer for human activities and work in the entertainment sector on the one hand, and a need for more residential development catering for the needs of people providing their workforce for this expanding industry. The linear trend statistics in Table 6 reveal highly significant growth trends for both demographic variables.

**Table 5**
Goodness of the linear fits to visitor and population statistics

| Fit Parameter | Visitors | Population |
|---|---|---|
| Slope ($b_1$) | 1,1828 | 19,0723 |
| Intercept ($b_0$) | -2,333 | -3766 |
| $r^2$ | 0,9657 | 0,9955 |

**Table 6**
Linear regression statistics with Student's t, degrees of freedom (df) and probability limits (p) for visitor and population statistics

| Linear Regression | Visitors | Population |
|---|---|---|
| t | 15,70 | 14,20 |
| df | (1, 24) | (1, 24) |
| $p$ | <.001 | <.001 |

**Figure 5**
Linear fits to visitor (top) and population (bottom) statistics for Greater Las Vegas

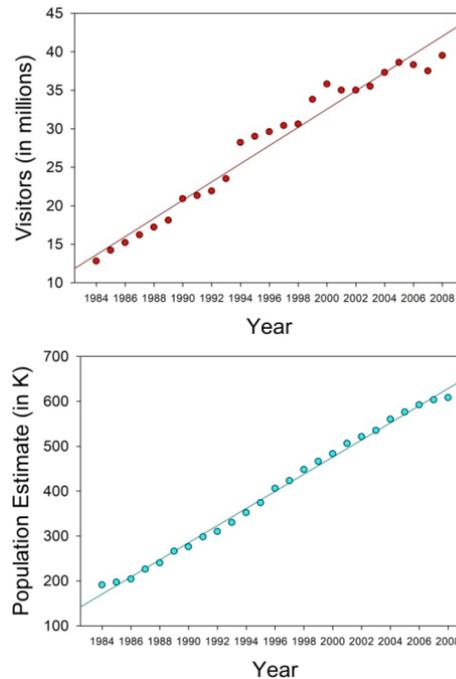

The subsequent analysis is concerned with the correlations between the QE distributions from image analysis and the distributions reflecting human impact data in terms of population and visitor growth over the years. Correlations are useful because, although a direct causality cannot be inferred, they may indicate a predictive relationship between variables. Significance and direction of correlations then can be further exploited for further interpretation of the image data in their wider context. To that effect we computed Pearson's correlation coefficient R, which mathematically determines statistical covariance. The probability p that the covariance of two observables is statistically significant is determined by the magnitude of the Pearson coefficient, which is directly linked to the strength of correlation, while its sign is directly linked to the direction of the covariance (positive or negative) of two variables. This analysis was performed on the paired distributions from the image analysis of the residential North and the population data, and on the paired distributions from the image analysis of Las Vegas City and the yearly visitor estimates. The correlations are shown graphically in Figure 6 for the SOM-QE data from the 25 images of the residential North as a function of the yearly population estimates (top) and the SOM-QE data from the 25 images of Las Vegas City as a function of the yearly number of visitors (bottom). The analysis signals statistically significant positive correlations between the paired variables in both cases. The correlation statistics, with Pearson coefficients R for a given comparison and the associated df and probability limits p, are shown in Table 7.

**Figure 6**

Correlations between SOM-QE and population (top) and visitor (bottom) statistics

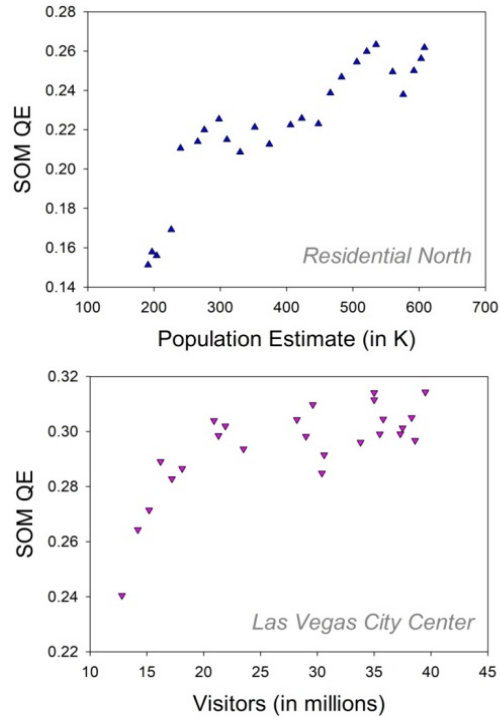

**Table 7**

Correlation statistics with Pearson coefficients R and the associated df and probability limits p

|  | SOM-QE Residential North *vs* Population | SOM-QE Las Vegas City *vs* Visitors |
|---|---|---|
| R | 15,70 | 14,20 |
| df | (1, 24) | (1, 24) |
| *p* | <.001 | <.001 |

Greater Las Vegas is one of the driest regions in the world (Frumkin et al , 2020). The particular problem space of considerable urban expansion and economic growth in such a region, and their impact on limited natural resources (Figure 1) therein, such as water, is highlighted by further statistics. The University of Nevada's Center for Gaming Research Annual Statistics (2023) show a steep increase in revenue from gaming and leisure activities in these years (1984-2008), while the water level measures for Lake Mead from the US Department of Interior's Bureau of Reclamation Hoover Dam Control Room Statistics (2023) show supply dwindling away during these same years. This is shown here in Figure 7 in terms of a significant negative correlation between economic growth in terms of a steep increase in revenue and water availability.

## 4. Discussion

SOM analysis of satellite images of the residential North of Greater Las Vegas and Las Vegas City Center consistently captures the structural changes in each ROI across the study years. These were characterized by a step-by-step reorganization of the City Center, with old casinos and hotel centers disappearing and replaced by more and larger new ones, and the progressive development of housing and infrastructure in desert regions that are now part of the residential North. The significant positive correlations between the QE distributions from the SOM-QE analysis of the satellite images and the statistical distributions on human impacts in terms of population growth and increasing visitor volume across the same years allow for a meaningful image interpretation in a wider sense. The human response to an increasingly larger offer of state-of-the-art entertainment in the City engendered an increasing need for residential development in desert regions to provide housing for an increasing population providing the necessary workforce. At the same time, the water supply to these arid regions dwindled away steeply. Greater Las Vegas being one of the most arid regions of the planet (Frumkin et al., 2020), linking the numerical model predictions statistically to numerical human impact data, as shown here, permits anticipating the scale of the effects of urbanization and directly related human impacts on this particular region.

**Figure 7**
Economic growth vs water availability (University of Nevada's Center for Gaming Research Annual Statistics and US Department of Interior's Bureau of Reclamation Hoover Dam Control Room statistics for the years 1984-2008)

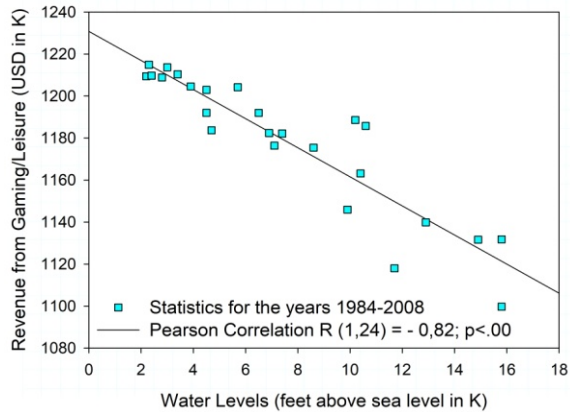

Such anticipation allows for prompt planning and implementation of measures for their mitigation, taking fully into account the specific geographic and economic context.

## 5. Conclusions

These data and observations illustrate how AI-assisted image analysis may be combined with the analysis of human impact statistics to enable an interpretation of image data in their direct societal context. The approach illustrates an explainable AI method for the rapid analysis of large data input streams by exploiting self-organized competitive learning for fully automated data analysis, in this study here pixel-based satellite imaging data, prior to further scrutiny or human decision making. The method is economic in computational resources and easily implemented, and can be used as a building block within step-by-step approaches to image analysis (e.g. Olsavszky et al, 2020), from pixels to image regions (e.g. Orheim and Lucchitta, 1990, Wandeto and Dresp-Langley, 2023). Current criteria for trustworthy and sustainable AI (HLEGAI, 2020, Radclyffe, Ribeiro, and Wortham, 2023) are satisfied for a wide range of potential applications, with limitations similar to those of any method of automated image data analysis. Prior hypotheses and additional analyses are required to convey meaning to the classification data.

## Acknowledgement

This work is sponsored by CNRS France and DEKUT Kenya.

## Ethical Statement

This study does not contain any studies with human or animal subjects performed by any of the authors.

## Conflict of Interest

The authors declare that they have no conflict of interest.

## Data Availability Statement

The data that support the findings of this study are available in the paper. The images analyzed here are available publically on Researchgate at the following page:
https://www.researchgate.net/publication/377577425_SatelliteImage_DataLVCityResidentialNorthtar.Copying the link into your browser opens the page that contains the .tar file for downloading.